\documentclass{easychair}

\usepackage{listings}
\lstdefinelanguage{json}{
    basicstyle=\ttfamily\small,
    breaklines=true,
}
\usepackage{booktabs}
\usepackage{orcidlink}
\usepackage{tcolorbox}
\usepackage{xcolor}
\definecolor{skyblue}{HTML}{5CE1E6} 
\usepackage{caption}
\usepackage{setspace}
\usepackage{hyperref}
\usepackage{minted}
\usepackage{pifont}
\usepackage{graphicx}
\usepackage{subcaption}
\usepackage[T1]{fontenc}
\usepackage{lmodern}
\usepackage{textcomp}
\begin{document}

\title{Accelerating Incident Response: A Hybrid Approach for Data Breach Reporting}

\author{
  Aurora Arrus\inst{1,2} \and
  Maria Di Gisi\inst{1,3}\orcidlink{0009-0003-5434-5426} \and
  Sara Lilli\inst{1,4}\orcidlink{0009-0000-2796-3067}\and 
  Marco Quadrini\inst{1,5}\orcidlink{0009-0002-4392-8186}
}

\institute{
  IMT School for Advanced Studies Lucca, 55100 Lucca, Italy \\
  \email{\{aurora.arrus, maria.digisi, sara.lilli, marco.quadrini\}@imtlucca.it}
  \and
  University of Cagliari, 09124 Cagliari, Italy\\
  \and
  University of Salerno, 84084 Fisciano, Italy \\
  \and
  Sant'Anna School of Advanced Studies, 56127 Pisa, Italy \\
  \and
  University of Camerino, 62032 Camerino, Italy \\
}

\authorrunning{Arrus, Di Gisi, Lilli, Quadrini}
\titlerunning{Accelerating Incident Response}

\maketitle

\begin{abstract}
The General Data Protection Regulation (GDPR) requires organisations to notify supervisory authorities of personal data breaches within 72 hours of discovery. Meeting this strict deadline is challenging because incident responders must manually translate low-level forensic artefacts such as malware traces, system-call logs, and network captures into the structured, legally framed information required by data-protection authorities. This gap between technical evidence and regulatory reporting often results in delays, incomplete notifications, and a high cognitive burden on analysts.
We propose a hybrid malware analysis pipeline that automates the extraction and organisation of breach-relevant information, with a particular focus on exfiltration-oriented Linux/ARM malware, which is rapidly increasing in prevalence due to the widespread adoption of IoT and embedded devices. The system combines static analysis to identify potential exfiltrators with dynamic analysis to reconstruct their behaviour. It employs a Large Language Model (LLM) constrained by a formal JSON schema aligned with the official Italian Garante Privacy notification form. The LLM transforms heterogeneous forensic artefacts into a structured, compliance-ready report that a human operator can rapidly validate.

\end{abstract}

\section{Introduction}
Personal data breaches create a critical operational risk for organisations that handle digital information. Modern regulatory frameworks intensify this risk by imposing strict transparency and accountability requirements. In the European Union, the General Data Protection Regulation (GDPR) requires controllers to notify the supervisory authority within 72 hours of discovering a breach and to provide a structured account of the incident, including the affected data and the mitigation measures adopted \cite{gdpr}. Other jurisdictions impose similar obligations \cite{hipaa,pipa_canada}, reinforcing a global shift toward the rapid and well-documented reporting of incidents. In practice, incident responders must interpret low-level forensic artifacts, such as malware binaries, system call traces, and packet captures, and convert them into the legally framed fields required by data protection authorities. 
This challenge becomes significantly harder when responders encounter Linux/ARM malware, whose prevalence continues to rise with the expansion of IoT and embedded devices \cite{iot_malware_growth}. During an investigation, analysts must determine whether a recovered binary performs data exfiltration, identify the resources it accesses, reconstruct the communication channels it uses, and present these findings in a form that aligns with regulatory templates. Existing malware-analysis pipelines, static, dynamic, or hybrid, support detection, family attribution, and threat hunting, but they rarely address regulatory reporting requirements. 
Recent applications of Large Language Models (LLMs) in security perform tasks such as log summarisation, triage, or explanation \cite{zhao2024apppoet,qian2025lamd}. However, they too generate unstructured narratives that still require extensive manual refinement before they can be made suitable for compliance workflows.
In this work, we adopt a different perspective, treating malware analysis as the first step in a workflow that ultimately accelerates GDPR breach notification. We focus on exfiltration-oriented Linux/ARM malware and combine static classification, selective dynamic analysis, and schema-constrained LLM reasoning to convert heterogeneous forensic artefacts into notification-ready content. 
This combination positions our work at the intersection of malware analysis, AI-assisted security automation, and legal-compliance support.
The rest of the paper is organised as follows: Section~\ref{chap:releated_works} reviews related work on malware analysis, exfiltration detection, and LLM-based security automation. Section~\ref{chap:methodology} describes the proposed pipeline, including dataset construction, static and dynamic analysis, and schema-constrained report generation. Section~\ref{chap:case_study} provides an end-to-end case study demonstrating how the system transforms a Linux/ARM malware sample into a breach-notification draft. Finally, Section~\ref{chap:conclusion} discusses limitations, outlines future work, and concludes the paper.
\section{Related Works}
\label{chap:releated_works}

Research on malware analysis, security automation, and privacy regulation spans several largely independent strands. 
Early work on static malware analysis examined how far one can understand malicious programs without executing them. These approaches translated binaries into intermediate structures such as control-flow graphs \cite{allen1970controlflow}, data-flow graphs \cite{kildall1973dataflow}, and abstract models \cite{cousot1977abstract} and reasoned about these representations to infer behavioural intent. Bergeron et al. \cite{bergeron2001static} demonstrated that reconstructed control-flow graphs, combined with model checking, could already capture characteristic malicious patterns. At the same time, Ding et al.\cite{ding2011feature} demonstrated that system-call sequence features extracted statically remained effective even under obfuscation, highlighting the resilience of semantic cues embedded in low-level code. Kinder et al.\cite{kinder2005detecting} extended these ideas by verifying behavioural specifications over abstracted program models, pushing static analysis toward more expressive semantic reasoning.

Over the past decade, this foundation has evolved toward richer and increasingly structural representations. Graph-based models now play a central role in capturing the internal logic of binaries. Recent work, such as that of MacDonald et al. \cite{macdonald2024binary}, illustrates this shift by combining call-graph topology with embeddings generated from decompiled code.


These developments underline that static analysis has become increasingly graph-centric, leveraging the rich topology of binary programs to infer behavioural properties.

While these works provide valuable insights and motivate our use of graph-based representations, they generally target broad detection or attribution tasks and rarely examine malware with respect to the specific behaviour of data exfiltration. 

On the other hand, dynamic analysis executes malware in controlled environments to capture runtime behaviour such as API calls, system interactions, and network traffic \cite{egele2012survey}. Numerous studies have shown that dynamic traces are highly informative for detection. 
For instance, Chanajitt et al.\cite{chanajitt2021combining} systematically compare models trained on static, dynamic, and combined features from Falcon Sandbox reports and PE metadata, and conclude that hybrid signals yield more robust detection. Industrial sandboxes, such as Cuckoo \cite{guarnieri2012cuckoo}, and commercial platforms integrate static metadata with behavioural logs to support triage and family attribution. However, these systems largely target x86/Windows binaries and focus on generic malware detection rather than on exfiltration-oriented Linux/ARM samples. 
In this scenario, data exfiltration detection has received substantial attention in network-centric security. Surveys such as Sabir et al. \cite{sabir2021machine} review machine-learning-based countermeasures and emphasise the difficulty of identifying stealthy exfiltration, given encryption, protocol mimicry, insider threats, and limited labelled datasets. 


Also, Large Language Models are increasingly used to support security operations, particularly for summarisation, triage, and incident reasoning. Surveys on AI-augmented Security Operations Centres (SOCs) describe LLM-driven workflows for alert analysis, timeline reconstruction, and automated report generation, showing that these models can significantly reduce manual overhead when interpreting heterogeneous security artefacts \cite{wang2024llm_security_survey}. Other works investigate Natural Language Processing (NLP) and LLM-based techniques for analysing logs, correlating indicators of compromise, and producing narrative incident summaries \cite{ogundairo2024nlp_incident_analysis,atzeni2025llm_siem,chernyshev2023llm_forensics}. 

While these systems demonstrate the value of LLMs for synthesising complex security data, none of them address the issue of transforming low-level forensic artefacts into the structured fields required for GDPR breach notifications. 

In parallel, regulators, standardisation bodies, and vendors have developed tools and guidelines to support breach notification under GDPR and related regimes. 
Commercial platforms such as OneTrust, RadarFirst, Securiti, and BreachRx, as well as other privacy-management or incident-response solutions, automate parts of the breach-management workflow by tracking incidents, mapping them to jurisdiction-specific notification rules, and managing deadlines and documentation \cite{onetrust_breach_response, radarfirst_breach, securiti_breach, breachrx_gdpr_guidelines}. These tools primarily operate at the policy and governance level, orchestrating processes, encoding legal requirements, and managing communication channels. However, they do not analyse low-level forensic artifacts, such as binaries, system-call traces, or PCAP files. As a result, organisations still rely on human experts to extract breach-relevant facts from technical evidence before they can use these platforms effectively.

\section{Methodology}
\label{chap:methodology}
\begin{figure}[ht]
    \centering
    \includegraphics[width=1\linewidth]{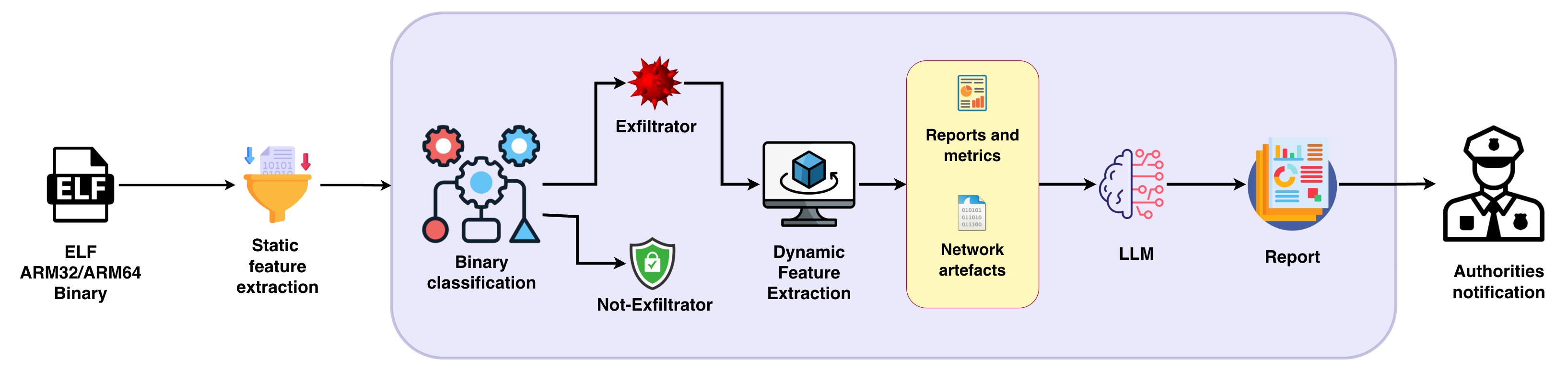}
    \caption{Pipeline}
    \label{fig:pipeline}
\end{figure}

This section details the proposed methodology for the LLM-assisted approach to GDPR data breach notification, leveraging insights from malware dynamic analysis. The complete process, illustrated in Figure \ref{fig:pipeline}, is organised into several interconnected phases, designed to transform raw technical forensic artifacts into a compliance-ready report.

The methodology begins with the ingestion of a suspect binary within the analysis pipeline. The first task is to determine the nature of the sample, specifically understanding whether it can be categorised as an exfiltrator malware and therefore warrants deeper investigation to identify the data it attempts to extract. This relies on a preliminary graph-based static analysis for feature extraction. From these graph-derived representations, a binary classification model determines if the sample exhibits exfiltration capabilities. Only samples identified as exfiltrators proceed to the next stage, where the binary is executed within a controlled emulation environment to capture behavioural indicators, such as system calls and API sequences, and generate network artifacts. These execution artifacts are subsequently fed into a Large Language Model (LLM), which, through a structured prompting mechanism, synthesises the technical findings into a human-readable report suitable for regulatory authorities.

\subsection{Dataset}
The dataset used in this study was constructed from samples retrieved via VirusTotal~\cite{virustotal}.
Based on the aim of this project, two distinct malware classes were collected. The first class consists of binaries returned by a query selecting all samples tagged by VirusTotal with MITRE ATT\&CK tactics~\cite{mitre-attack} corresponding to data access and exfiltration behaviours. Specifically, the query included the categories listed in Table~\ref{tab:attack_categories}.

\begin{table}[ht]
\centering
\caption{ATT\&CK categories used to define the exfiltrator class in the dataset.}
\label{tab:attack_categories}
\begin{tabular}{lc}
\toprule
\textbf{ATT\&CK ID} & \textbf{Description} \\
\midrule
TA0006 & Credential Access \\
TA0010 & Exfiltration \\
TA0009 & Collection of sensitive data \\
TA0100 & ICS data collection \\
TA0035 & Mobile data collection \\
TA0031 & Credential access on mobile \\
TA0036 & Mobile data exfiltration \\
\bottomrule
\end{tabular}
\end{table}

These categories were adopted as assigned by the VirusTotal tagging system, which internally maps behaviours to MITRE ATT\&CK.
The second class includes malware samples not labelled under any of the categories listed above, and then treated as a non-exfiltrator category. All collected samples were filtered to retain only \textbf{ARM32} and \textbf{ARM64} Linux ELF binaries.
This design choice reflects the objective of targeting architectures representative of mobile and IoT ecosystems rather than conventional x86\&x86\_64 desktop or server environments. Although the proposed dynamic-analysis framework supports more architectures and their combinations with ABIs, endianness and libraries, the static feature-extraction pipeline requires architecture-specific handling, because the graph-based representations employed depend on specific register sets and instruction semantics. These features vary significantly across CPU architectures. To maintain methodological consistency and ensure a controlled experimental setup, the dataset was restricted to ARM binaries. For the experimental evaluation, a balanced subset of \textbf{512 exfiltrators} and \textbf{512 non-exfiltrators} was sampled. 
Table~\ref{tab:arch-distribution} summarises the dataset distributions over architecture, endianness, bit-addresses and libc implementation.

\begin{table}[ht]
\centering
\caption{Distribution of dataset architecture for exfiltrator and non-exfiltrator samples.}
\label{tab:arch-distribution}
\begin{tabular}{llcc}
\toprule
\textbf{Property} & \textbf{Value} & \textbf{Exfiltrators (512)} & \textbf{Non-exfiltrators (512)} \\
\midrule
Architecture & ARM & 512 & 512 \\
\midrule
Bit-address & 32 & 62 & 450 \\
Bit-address & 64 & 119 & 393 \\
\bottomrule
\end{tabular}
\end{table}

\subsection{Static Feature Extraction}
Static analysis consists in studying the binary code without executing it, usually through pattern matching, signatures, or control-flow analysis~\cite{ref_article5}. In this study, the goal of the first phase is, in fact, to transform each raw binary sample into a set of structured graph representations that capture meaningful semantic information without executing the code. Concretely, our static pipeline produces three complementary artefacts: an import graph that summarises external dependencies, a function call graph (FCG) enriched with per–function features, and an interprocedural control–flow graph (ICFG) at basic–block granularity. All graphs are constructed programmatically and exported both as NetworkX GraphML files and as JSON summaries to facilitate downstream machine–learning experiments.

\paragraph{Function Call Graph}
\label{paragraph:fcg}
The function–level structure of the program is captured by a separated module that uses the \texttt{angr} framework to recover an approximate control–flow graph via \texttt{CFGFast} with libraries disabled (\texttt{auto\_load\_libs=False}). The script enumerates all discovered functions and constructs a directed FCG in which nodes correspond to functions and edges to statically inferred call relations. To identify calls, the implementation scans all basic blocks in each function and disassembles their instructions using angr’s Capstone integration when available, falling back to the standalone Capstone Python bindings otherwise. 
It specifically looks for ARM call–like instructions such as \texttt{bl} and \texttt{blx}; for each such instruction, it attempts to resolve the call target either via an immediate operand (when present) or by parsing the textual operand string and interpreting hexadecimal constants as candidate addresses. 
If the resolved address does not exactly match the entry point of a known function, the script performs a best–effort search over the function list and assigns the edge to the unique function whose address range contains the target.

Each FCG node is annotated with a set of per–function static features that summarise its local structure computing, for every function, the number of basic blocks, the total number of bytes across its blocks, an estimate of the instruction count, and the average and standard deviation of Shannon entropy over basic–block byte sequences. 
In parallel, it collects control–flow–related statistics by analysing the disassembled instruction stream: the number of direct calls, the number of indirect calls (e.g., \texttt{blr}, \texttt{bx}), and the number of supervisor calls (\texttt{svc}/\texttt{swi}), which approximate syscall usage. These quantities are stored as numeric node attributes in the FCG. From a classification perspective, these features encode both structural properties (e.g., size, fragmentation, branching density) and behavioural hints (e.g., frequent indirect calls or syscalls), which are known to be discriminative for distinguishing benign and malicious code. The final JSON export includes the number of nodes and edges, as well as the full dictionary of per–function attributes.

An example of the generated JSON for the Function Call Graph is shown below, illustrating the top-level metadata and the feature dictionary associated with a single function node.

\begin{lstlisting}
{
  "binary": "dataset/malware.elf",
  "n_nodes": 610,
  "n_edges": 517,
  "nodes": {
    "10485984": {
      "addr": 10485984.0,
      "name": "sub_a000e0",
      "n_blocks": 4.0,
      "total_bytes": 28.0,
      "avg_block_entropy": 2.6045739585136225,
      "std_block_entropy": 0.6223994673127933,
      "bl_count": 4.0,
      "bl_indirect_count": 0.0,
      "svc_count": 0.0,
      "avg_bytes_per_block": 6.99999999825,
      "instr_count_est": 7.0
    }
  }
}
\end{lstlisting}

\paragraph{Interprocedural Control Flow Graph}
While the FCG operates at function granularity, the ICFG produced by this module refines the analysis to the level of individual basic blocks and explicit control–flow transfers. Using the same \texttt{angr} configuration as above, the script constructs a normalised \texttt{CFGFast} and then iterates over all nodes in the underlying control–flow graph. For each node, it obtains the corresponding basic block, extracts the raw bytes, and computes their Shannon entropy using a 256–bin byte histogram. 
The ICFG node set is then defined by the set of basic–block addresses; each node is annotated with its start address, size in bytes (when available), and entropy value, which provides a coarse measure of randomness and potential packing. Edges in the ICFG are taken directly from the CFG: for every pair of source and destination CFG nodes, a directed edge is inserted between the corresponding basic–block nodes in the NetworkX graph. This representation captures the detailed control–flow topology of the binary, including interprocedural transitions, and is therefore well-suited for graph–based learning methods that operate on fine–grained inputs. 
As in the previous cases, the script exports both a GraphML file encoding the graph and a JSON file containing the set of nodes and their attributes.

The ICFG is exported as a JSON structure where each basic block is represented as a node enriched with size and entropy information, as shown in the excerpt below.

\begin{lstlisting}[basicstyle=\ttfamily\small,breaklines]
{
  "binary": "dataset/0c2534548bc8e4fb64425d7d0589b54b078767cf6f1bec6b38dfaedf308f85a1",
  "n_nodes": 8063,
  "nodes": {
    "33168": {
      "addr": 33168,
      "size": 44,
      "entropy": 4.047182403014169
    },
    "32916": {
      "addr": 32916,
      "size": 16,
      "entropy": 3.875
    }
  }
}
\end{lstlisting}

This multi–graph representation with these two artefacts form the basis for the learning components of our pipeline, where graph–based models exploit both topology and node attributes to discriminate between exfiltrator and non-exfiltrator malicious binaries.

\subsection{Binary Classification}
\label{subsec:binary-classification}

The objective of this stage is to determine whether a binary sample exhibits structural and behavioural characteristics typically associated with data-exfiltrating malware. The prediction relies exclusively on static information extracted from the Function Call Graphs (FCGs) described in Section~\ref{paragraph:fcg}, which summarise the program’s internal organisation and inter-procedural control-flow relationships while avoiding the need for execution.

For each sample, a numerical representation was constructed that integrates global topological descriptors, aggregated function-level attributes, and embedding-based features. The topological descriptors quantify the high-level organisation of the call graph by measuring its size, connectivity, degree distributions, component structure, and the relative prevalence of entry and exit nodes. 


This representation was enriched with aggregated node-level attributes reflecting the internal structure of the program’s functions. These measurements include basic-block composition, code size, instruction density, and entropy-based indicators that frequently reveal code packing, encryption, or other obfuscation practices, which attackers commonly employ to conceal exfiltration logic. Additional behavioural cues derived from call patterns provide further context, particularly when functions repeatedly interact with privileged services or system-level primitives associated with file or network operations.

To capture latent structural regularities that handcrafted statistics may overlook, we encode each FCG using a graph embedding process. This mapping places every function node into a low-dimensional vector space that reflects its role within the broader call topology. The embeddings preserve higher-order proximity relationships, enabling the classifier to exploit subtle interaction patterns and relational information not directly visible from explicit feature engineering. After aggregating the node embeddings at the graph level, we obtain a complementary feature component that enriches the overall representation. Concatenating all descriptors yields a fixed, 76-dimensional vector for each binary, combining coarse structural cues, fine-grained code attributes, and relational information.

To distinguish exfiltrators from non-exfiltrators, a Random Forest classifier was trained on a dataset that comprises 1024 labelled Linux ELF malware samples, evenly balanced between exfiltrators and non-exfiltrators. 
The dataset includes samples from multiple malware families to prevent family-dependent biases and ensure that the classifier generalises across diverse structural patterns. 
The dataset was split into 75\% for the training set and 25\% for the test set.

Table~\ref{tab:classification_metrics} reports the model’s performance on the test set. The classifier achieves high accuracy and balanced precision–recall values, confirming that static structural information alone is sufficiently expressive to discriminate exfiltration-oriented malware with high reliability. 

\begin{table}[ht]
\centering
\caption{Performance metrics of the binary classifier on the 25\% held-out test set.}
\label{tab:classification_metrics}
\begin{tabular}{lc}
\toprule
\textbf{Metric} & \textbf{Value} \\
\midrule
Accuracy & 0.967 \\
Precision & 0.958 \\
Recall & 0.972 \\
F1-score & 0.965 \\
ROC--AUC & 0.983 \\
\end{tabular}
\end{table}

\subsection{Dynamic Feature Extraction}


While the static phase provides a structural and semantic characterisation of each sample, dynamic analysis is only triggered for binaries that the classifier labels as potential \emph{exfiltrators}. 
Dynamic analysis executes the program in a monitored environment and observes its runtime behaviour. This approach is more informative with respect to static analysis, because it can capture live API calls, system calls, arguments, and network activity~\cite{ref_article6,ref_article7}.
The goal of this stage is no longer to decide whether the sample is malicious, but to reconstruct \emph{what data} it attempts to access and transmit, and through which channels, in order to support GDPR breach notification and impact assessment. To this end, we rely on \textbf{Emulix}, a containerised reverse engineering and malware analysis platform specifically designed for heterogeneous Linux malware.  Emulix executes ELF binaries inside an isolated Docker sandbox, with internet connectivity disabled and all outbound traffic transparently redirected to a dedicated network container running FakeNet-NG. 
This design ensures the possibility of reproducing tests while still allowing for realistic observation of network behaviour: the malware "believes" it is communicating with external services, whereas all traffic is mediated and recorded by FakeNet-NG.

\paragraph{Execution environment and architecture support.}
Unlike traditional full-system virtualisation, Emulix does not boot a complete guest operating system. 
Instead, it instantiates a minimal userspace that understands the underlying architecture, obtained from prebuilt Buildroot root file systems. 
Metadata parsed from the ELF headers (architecture, ABI, endianness, word size, and C library) is used to select a compatible root file system image, which is mounted inside the sandbox container before execution. 
Table~\ref{tab:rootfs-configs} summarises the configurations currently supported in our prototype for the GDPR notification, focusing on ARM-based platforms.

\begin{table}[h]
\centering
\caption{Root file system combinations currently supported.}
\label{tab:rootfs-configs}
\begin{tabular}{llll}
\toprule
\textbf{Architecture} & \textbf{Endianness} & \textbf{Word size} & \textbf{C library} \\
\midrule
ARM       & little (lsb) & 32 & glibc   \\
ARMhf     & little (lsb) & 32 & glibc   \\
ARM       & little (lsb) & 32 & uclibc  \\
ARM       & little (lsb) & 64 & glibc   \\
ARM       & little (lsb) & 64 & uclibc  \\
ARM       & big (msb)    & 32 & uclibc  \\
\bottomrule
\end{tabular}
\end{table}

Although Emulix and the underlying raw emulation framework (Qiling) are capable of handling additional architectures such as x86 and MIPS, extending support to new platforms in our pipeline requires implementing the corresponding static feature extraction for that ISA so that dynamic traces can be meaningfully correlated with the graph-based representations used in the previous phase.

\paragraph{Emulation workflow.}
Given a flagged exfiltrator binary and its metadata, Emulix synthesises the execution environment by mounting the selected root file system inside the sandbox container and launching the sample via the Qiling framework. 
Execution proceeds on top of the Unicorn CPU emulator, and each system call or high-level API call is hooked by a custom handler invoked prior to the actual emulation of the call. 
Within this handler, Emulix logs the invoked function, normalised arguments (e.g., file descriptors resolved to paths, flags mapped to symbolic constants, socket families converted to protocol names), and the CPU register state at the call site. 
Optionally, full memory dumps of the process can be captured at configurable points for forensic use.

Network traffic is never sent directly to the Internet. 
All outbound connections are transparently redirected to the FakeNet-NG container, which simulates common network services, generates protocol-consistent responses, and records complete packet captures in PCAP format. 
This allows us to safely observe exfiltration attempts, reconstruct destination endpoints, and inspect payload contents without exposing real external infrastructure.

At the end of execution, or when a configurable timeout is reached, the sandbox is torn down, and a collection of structured artefacts is exported to the host. 
Resource usage and runtime are bounded via container-level controls, making the dynamic phase practical even when multiple samples must be analysed in parallel.

\paragraph{Dynamic artefacts used in the pipeline.}
For each analysed binary, Emulix produces several outputs. 
In our pipeline we focus on those that are most informative for reconstructing exfiltration behaviour and feeding the downstream LLM:

\begin{itemize}
  \item \textbf{Emulation log (\texttt{emulation\_results.txt}):} a fine-grained execution trace with one entry per system call or API call. 
  Each record includes the invoked function, its normalised arguments, and the CPU register snapshot at the call site. 
  This file constitutes the primary behavioural timeline of the sample and is the main source for understanding which resources (files, sockets, descriptors) were accessed.

  \item \textbf{Qiling runtime output (\texttt{qiling\_output.txt}):} a low-level report emitted by the Qiling engine, including information about loaded libraries, memory mappings, relocation symbols, and syscall return values. 
  Although primarily intended for debugging and validation of the emulation, it can also support the interpretation of ambiguous behaviours observed in the higher-level logs.

  \item \textbf{Network artefacts (\texttt{packets\_*.pcap}):} full packet captures of all traffic routed through FakeNet-NG. 
  These PCAP files are used to inspect exfiltrated payloads, identify application-layer protocols, and correlate network transmissions with the resources accessed in \texttt{emulation\_results.txt}.

  \item \textbf{Execution flow traces (\texttt{ip\_trace.log} and \texttt{ip\_flow\_path.csv}):} chronological sequences of program-counter (RIP/PC) addresses recorded at each hooked event. 
  These traces describe how the binary traverses its code regions during emulation, highlighting execution hotspots, unpacking stubs, tight loops, and control-flow patterns associated with exfiltration routines.
\end{itemize}

In the context of this work, these dynamic artefacts are not used to train another classifier, but are instead consumed by the Large Language Model to reconstruct, in natural language, a detailed account of the incident. 

\subsection{Large Language Model Integration}
The final stage of the proposed pipeline integrates a Large Language Model (LLM) to bridge the gap between raw technical evidence and the regulatory requirements of the Italian Data Protection Authority (Garante per la Protezione dei Dati Personali). The system does not autonomously submit the notification; instead, it acts as a compliance assistant, producing structured and human-readable technical content to support operators in completing the official form. This significantly reduces the time and specialised expertise required to meet the 72-hour deadline following a breach.
The LLM, specifically Gemini 2.5 Pro\footnote{\href{https://docs.cloud.google.com/vertex-ai/generative-ai/docs/models/gemini/2-5-pro?hl=it}{Gemini 2.5 Pro documentation}}, accessed via the Vertex AI API using the \texttt{google.generativeai}\footnote{\href{https://pypi.org/project/google-generativeai/0.8.4/}{https://pypi.org/project/google-generativeai}} library, is configured with a System Instruction Prompt to interpret raw technical artifacts from dynamic analysis according to a strict JSON schema that mirrors the official Garante Privacy notification form (Sections F, G, H)\footnote{\href{https://servizi.gpdp.it/databreach/resource/1629905132000/DB_Istruzioni}{Fac-simile model for Italian data breach notification}}. This ensures that each extracted detail corresponds to the exact field required in the notification.

The report generation process proceeds in two main phases to handle large files and potential context limitations. 
\\
In Phase 1, smaller but crucial logs (\texttt{emulation\_results.txt}, \texttt{qiling\_output.txt}, \texttt{ip\_trace.txt}, \texttt{ip\_flow\_path.csv}) are analysed against the JSON schema, producing a preliminary structured report. 
\\
Phase 2 introduces the larger \texttt{packets\_*.pcap} file, enabling the LLM to enrich and refine the initial report by filling in additional relevant information. Finally, the structured JSON is transformed into a PDF using the ReportLab library\footnote{\href{https://pypi.org/project/reportlab/}{reportlab library}}, producing a human-readable document where each section of the official notification is paired with the corresponding extracted content. 

This integrated pipeline enables human operators to efficiently review, verify, and submit the necessary information, thereby bridging the gap between technical forensic analysis and regulatory compliance, while significantly reducing the time required to prepare the notification.

\subsection{GDPR Compliance}

From a legal standpoint, the system proposed in this work has been designed to operate in full alignment with the regulatory framework established by the GDPR, with specific reference to data breach notification obligations. Article 33 GDPR requires data controllers to notify the competent supervisory authority of any personal data breach without undue delay and, where feasible, within 72 hours of becoming aware of the incident. The notification must include a structured description of the breach, specifying, among others, the categories and approximate number of data subjects concerned, the categories and volume of personal data involved, the likely consequences of the breach and the technical and organisational measures adopted or proposed to mitigate its adverse effects \cite{borgesius2023gdpr}. 

In practice, incident response teams face significant operational difficulties in meeting these requirements, as the evidence generated during malware analysis and network monitoring is typically expressed through highly technical artefacts, such as execution traces, system logs, packet captures and binary analysis reports.

The system described in this paper directly addresses this issue by introducing an automated mechanism for converting technical evidence into structured, legally relevant information that aligns with the official breach notification model adopted by the Italian Data Protection Authority. 

The system's role is strictly limited to decision support, documentation, and information synthesis. This design choice ensures compliance with the GDPR principles of accountability and human oversight, and excludes any form of automated decision-making within the meaning of Article 22 of the GDPR.
Regarding the lawfulness of processing, the system is configured exclusively to support cybersecurity operations, incident response, and regulatory compliance. Consequently, the legal basis for the processing is twofold: (i) Article 6 (1)(f) GDPR, which recognises the protection of networks and information systems as a legitimate interest pursued by the controller; (ii) Article 6(1)(c) GDPR, insofar as the tool supports compliance with mandatory legal obligations relating to the notification and documentation of personal data breaches. 
These legal bases are further reinforced by Recitals 46, 49, and 76 of the GDPR, which expressly acknowledge cybersecurity as a fundamental and overriding interest.
Moreover, personal data are processed solely for the purpose of identifying compro
mised information and assessing the risk to data subjects, in accordance with the principles of data minimisation, purpose limitation and proportionality.

From a compliance perspective, the pipeline contributes not only to the timeliness of breach notifications but also to their substantive quality. By enabling structured, traceable and evidence-based reporting, the system strengthens the controller’s ability to demonstrate compliance under the accountability principle (Article 5(2) GDPR), while also reducing the risk of human error.


\section{Case Study}
\label{chap:case_study}
This section presents an end-to-end case study of an exfiltrating malware sample.
The sample has been chosen randomly among all the samples labelled as exfiltrators that presented a registry connection attempt, to understand the full potential of the analysis pipeline.
The sample in question is identified by the corresponding cryptographic hash computed on the binary itself, namely \texttt{a90c09ca384feaa4b7e111549bdafddbc778314257606ef61a6822c1cedb25cf}.
  
\begin{figure}[ht]
    \centering    \includegraphics[width=0.5\linewidth]{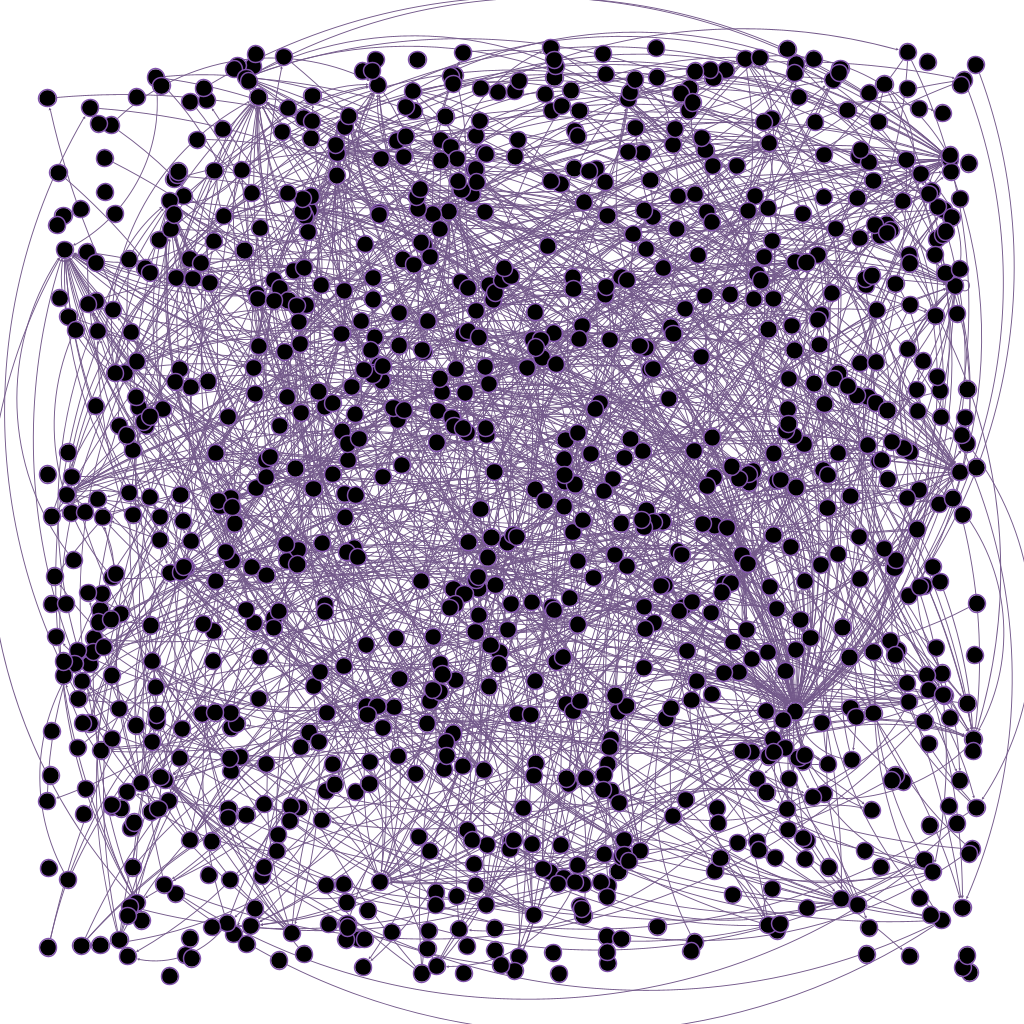}
    \caption{Control Flow Graph}
    \label{fig:fcg}
\end{figure}

\paragraph{Static Analysis.} The static analysis pipeline produced the import graph, ICFG, and Call Graph representations of the sample. The call graph (\texttt{*\_fcg.graphml}) enables a structural examination of the program's functional decomposition, with a total of {\tt 971} nodes and {\tt 1058} directed edges. The graph has an average out-degree of approximately $1.09$: this indacates that the graph is only lightly connected: on average, each function calls just about one other function.  
A single giant weakly connected component comprises $452$ functions, while the remaining functions are split between $4$ small components of size $2$ and $511$ isolated nodes with no incoming or outgoing calls.  
This indicates that the executable contains a dense behavioural core surrounded by a large number of helper or rarely used routines that play a much more limited or isolated role. The function call graph extracted is shown in Figure~\ref{fig:fcg}.

\begin{figure}[ht]
    \centering    \includegraphics[width=0.5\linewidth]{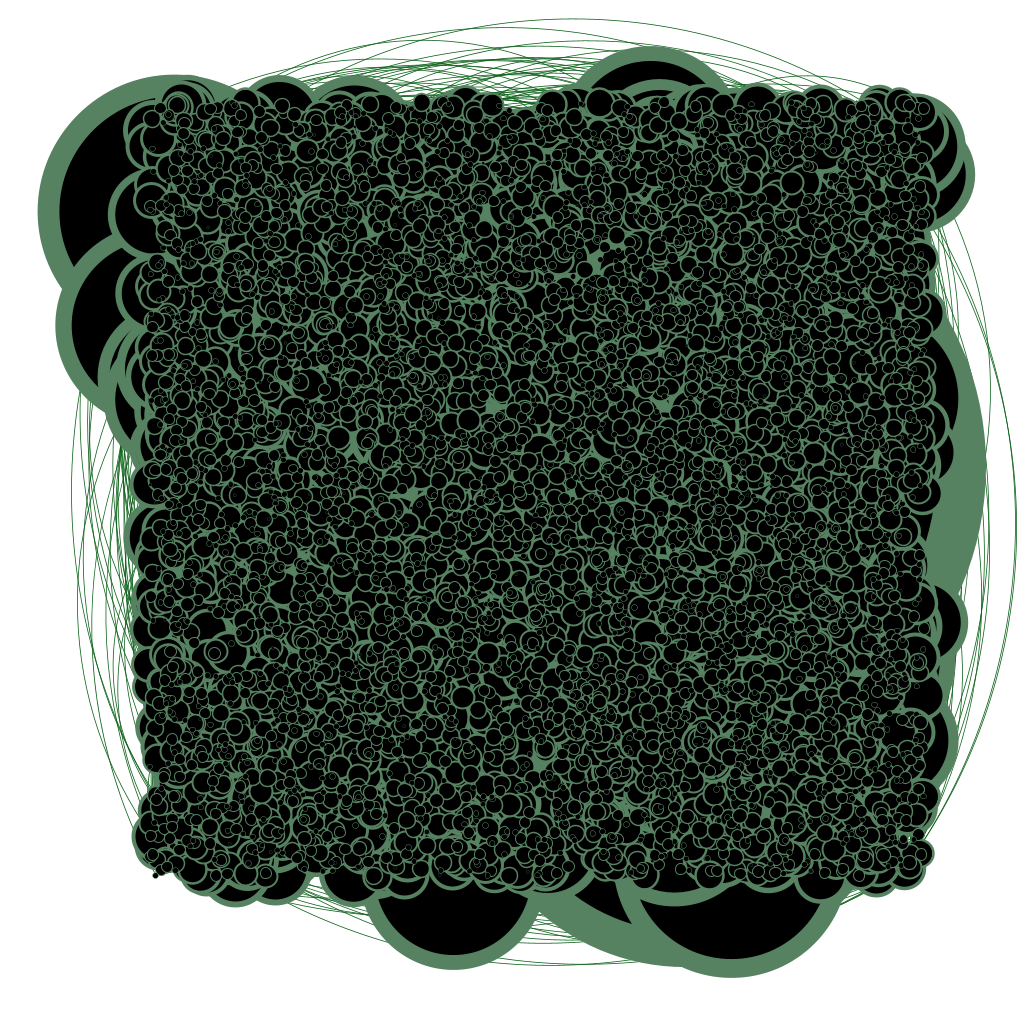}
    \caption{Interprocedural Control-Flow Graph}
    \label{fig:icfg}
\end{figure}

The interprocedural control-flow graph (ICFG) (\texttt{*\_icfg.graphml}) provides a finer-grained view of the program’s internal execution structure.  
The extracted graph contains {\tt 6511} basic-block nodes and {\tt 11893} directed edges, yielding an average out-degree of approximately $1.83$, which reflects a moderately connected control-flow topology.  
A single dominant weakly connected component contains more than {\tt 6133} blocks, indicating that the majority of the program participates in one large, contiguous execution region.  
The remaining components consist of small clusters of size $6$–$13$, together with {\tt 198} isolated blocks, which likely correspond to error paths, alignment padding, or compiler-generated stubs.  
Overall, the ICFG reveals that the binary is not monolithic but composed of a sizable execution core enriched with numerous side paths and secondary behaviours, consistent with malware that contains scanning loops, conditional branching logic, and exfiltration routines integrated into a broader control-flow backbone.
The extracted Interprocedural Control-Flow Graph is shown in Figure~\ref{fig:icfg} 

\paragraph{Classification.} The trained Random Forest model was placed in evaluation mode and confirmed that the binary was an exfiltrator (label 1).

\paragraph{Dynamic Analysis.} The dynamic execution of the sample on the ARM emulation environment offers a description of behaviours relevant for reporting.
The syscall trace shows initial reconnaissance activity (filesystem probing and process setup), followed by a renaming operation that replaces the binary’s pathname with \texttt{/usr/dvrHelper}, a persistence and masquerading technique commonly used on embedded Linux systems.  
The network monitor records {\tt 824} packets and {\tt 577} flows, with outbound DNS queries toward multiple external resolvers and a set of characteristic UDP/TCP exchanges.  
Among these, DNS traffic to external authoritative servers on port~53 indicates the presence of a DNS-based communication channel, while a small number of TCP flows to remote IP addresses demonstrates the establishment of direct outbound connections.  
In addition, a large volume of short-lived TCP flows targeting port 23 across more than 100 unique external IP addresses reveals a systematic scanning pattern consistent with the propagation phase of IoT malware.  
These observations collectively provide the behavioural footprint necessary to understand which data the sample tried to exfiltrate, and how. The detailed semantic interpretation of the attack (exfiltration channel, botnet role, C2 infrastructure, propagation strategy) is deferred to the LLM-generated report.

\begin{figure}[h!]
    \centering
    \begin{subfigure}{0.3\linewidth}
        \centering
        \includegraphics[width=\linewidth]{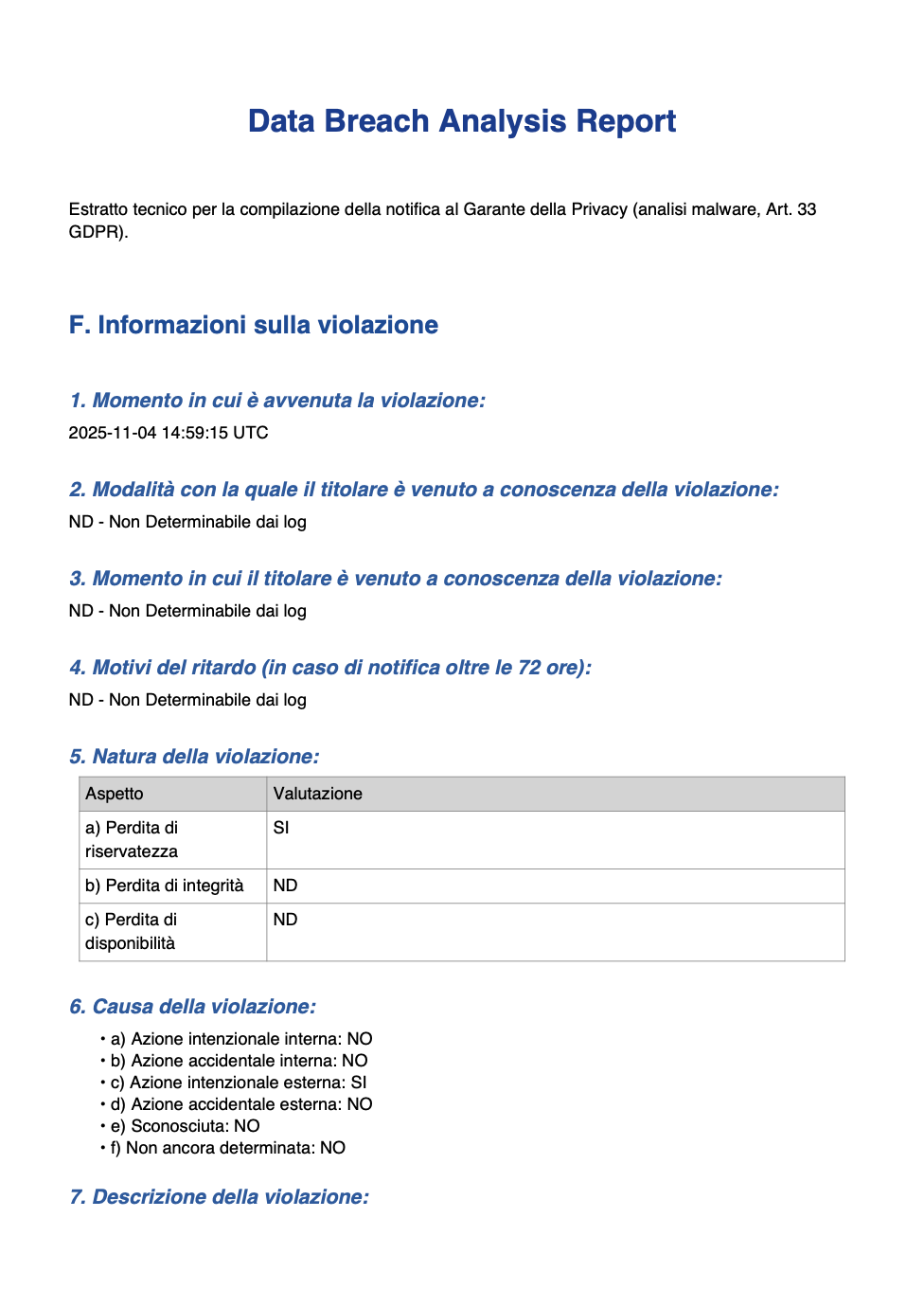}
    \end{subfigure}
    \hfill
    \begin{subfigure}{0.3\linewidth}
        \centering
        \includegraphics[width=\linewidth]{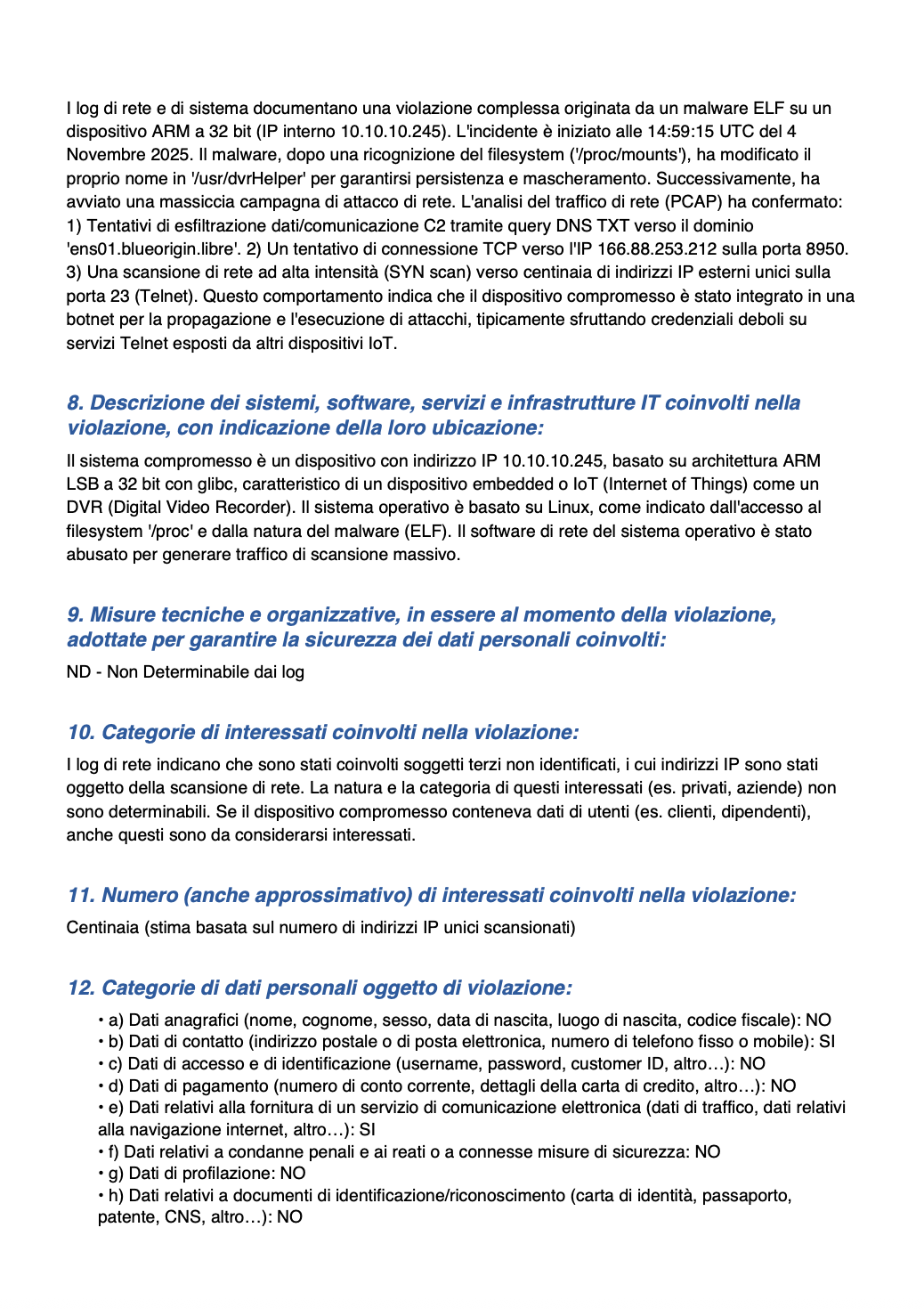}
    \end{subfigure}
    \hfill
    \begin{subfigure}{0.3\linewidth}
        \centering
        \includegraphics[width=\linewidth]{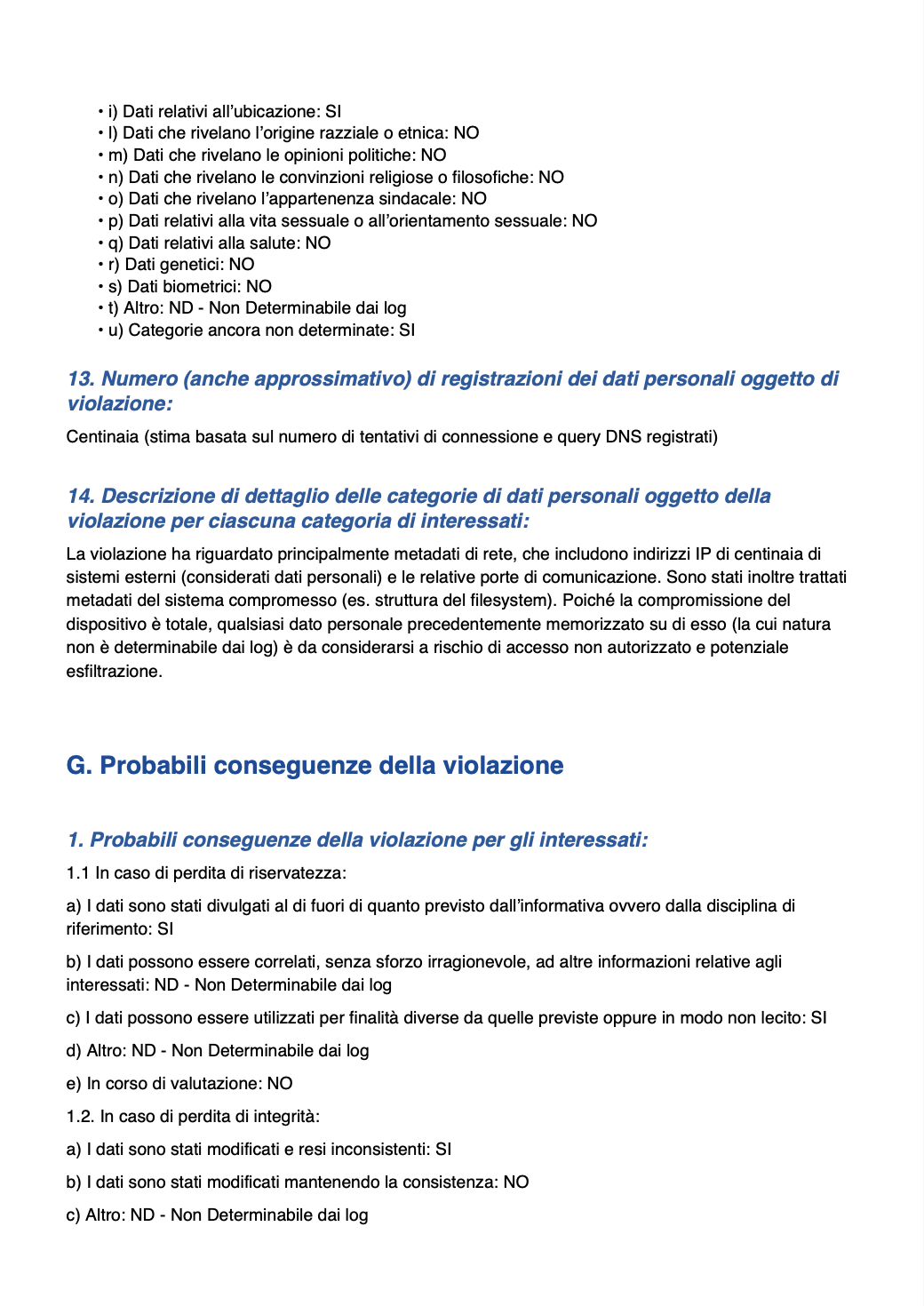}
    \end{subfigure}

    \caption{Partial Output of the LLM-Generated Structured Report. \\ \textit{Note: the contents are presented in Italian to faithfully replicate the mandatory section headings of the Italian Garante Privacy official notification model, which serves as the target schema for our compliance engine.}}
    \label{fig:report}
\end{figure}

\paragraph{Report Generation.} The LLM was invoked to extract crucial information from the raw data and populate the JSON conforming to the schema defined based on Sections F, G, and H of the Italian Garante Privacy notification model. The objective was to translate evidence such as IP addresses, accessed files, and protocols used into structured, narrative descriptions for the compliance officer. Figure \ref{fig:report} shows a partial and representative example of the output generated by the LLM, specifically tailored to the legal requirements of the Italian Data Protection Authority. The fields displayed in Italian directly mirror the mandatory sections of the official national notification form, thus validating the LLM's ability to produce a highly specific, compliance-ready artifact.

As demonstrated, the integration of the LLM enables the transformation of an execution trace and a PCAP into a structured artifact, ready for human review, thereby significantly reducing the burden and time required to prepare the official report.
\section{Discussion and Conclusion}
\label{chap:conclusion}

Our work successfully addresses the critical bottleneck in GDPR compliance: the manual translation of low-level forensic data into high-level, structured legal reporting. The integration of a Random Forest classifier utilising graph-based static features enables the rapid pre-filtering of ELF ARM binaries, demonstrating high discriminatory power (e.g., a $0.983$ ROC-AUC score) against exfiltrators. This static filtering step significantly reduces the operational load on the dynamic analysis engine, ensuring that detailed, time-consuming emulation is only performed on high-risk samples. Most importantly, the LLM-assisted reporting module, constrained by the Garante Privacy JSON schema, successfully transforms complex artifacts (such as execution traces and PCAPs) into a compliance-ready report. This capability is pivotal, turning the mandatory 72-hour notification window from a high-stress compliance risk into a semi-automated, manageable process.

Despite the demonstrated effectiveness, the pipeline contains inherent limitations that guide future research. Firstly, the static analysis component, while highly accurate, remains vulnerable to sophisticated obfuscation and dynamic unpacking techniques, potentially leading to false negatives against advanced persistent threats. Secondly, the dynamic analysis phase, operating within the Emulix environment, is susceptible to sandbox evasion techniques, where malware may detect the virtualised environment and withhold its exfiltration payload. Furthermore, while the structural constraints of the JSON schema mitigate the risk of LLM hallucination, the generated report still functions as a compliance aid and is not a fully autonomous notification, necessitating a final human review for absolute legal certainty.

In conclusion, we have presented an innovative, hybrid malware analysis pipeline that effectively tackles the time-sensitive challenge of data breach notification under the GDPR. Our approach combines an efficient static classifier for ARM ELF binaries with focused dynamic analysis and the structured reporting capabilities of the Gemini 2.5 Pro LLM. This unique combination not only quickly identifies exfiltration-oriented malware but also automatically generates a legally compliant notification artifact. By successfully bridging the divide between low-level technical evidence and stringent regulatory requirements, our system significantly reduces the manual preparation burden and provides crucial automated support for incident response teams striving to meet the strict 72-hour reporting deadline. 

\newenvironment{acknowledgements}{%
  \section*{Acknowledgements} 
}{\par}

\begin{acknowledgements}
This work was partially supported by project SERICS (PE00000014) under the MUR National Recovery and Resilience Plan funded by the European Union - NextGenerationEU.
\end{acknowledgements}

\section*{Declaration on Generative AI}
During the preparation of this work, the authors used ChatGPT, Grammarly in order to: Grammar and spelling check, Paraphrase and reword. After using these tools, the authors reviewed and edited the content as needed and took full responsibility for the publication’s content.

\bibliographystyle{unsrt}
\bibliography{references}

\end{document}